\documentclass[useAMS,usenatbib]{mn2e}

\usepackage[dvips]{graphicx}
\usepackage{natbib}

\title[The dust velocity near the Sun in 2006] {The velocity of the
dust near the Sun during the Solar Eclipse of March 29, 2006 and 
sungrazing comets}

\author[L. I. Shestakova et al.]{L. I. Shestakova$^{1}$, A. Chalabaev$^{2}$, 
B. I. Demchenko$^{1}$, and F. K. Rspaev$^{1}$
\thanks{E-mail: shest@aphi.kz (LIS), Almas.Chalabaev@obs.ujf-grenoble.fr (AC)} 
\footnotemark[1]\\
$^{1}$Kazakh National Space Agency, Fesenkov Astrophysical Institute,
Kamenskoie Plato, Observatory 23, Almaty, 050020, Kazakhstan\\
$^{2}$Laboratoire d'Astrophysique de Grenoble, UMR 5571, CNRS, Universit\'{e} 
Joseph-Fourier, BP 53X, Grenoble, CEDEX 09, France} 

\begin{document}

\date{Accepted xx. Received 2010 March 8; in original form 2010 xx xx}

\pagerange{\pageref{firstpage}--\pageref{lastpage}} \pubyear{200x}

\maketitle

\label{firstpage}

\begin{abstract} The measurements of the Doppler shifts of the
Fraunhofer lines, scattered by the dust grains in the solar F-corona,
provides the insight on the velocity field of the dust and hence on
its origin.  We report on such measurements obtained during the total
eclipse of March 29, 2006.  We used a Fabry-P\'{e}rot interferometer
with the FOV of $5.9\degr$ and the spectral resolution of about 5000
to record Fraunhofer spectral lines scattered by the dust of the
F-Corona.  The spectral region was centered on the \mbox{Mg\,{\sc i}}
$5172.69 \rmn{\AA}$ line.  The measured line-on-sight velocities with
the amplitude in the range from -10 to 10~km$\cdot \rmn{s^{-1}}$ show
that during our observations the dust grains were on the orbit with a
retrograde motion in a plane nearly perpendicular to the ecliptics.
This indicates their cometary origin.  Indeed, at the end of March,
2006, SOHO recorded several sungrazing comets with the orbital
elements close to what was deduced from our measurements.  We conclude
that the contribution of comets to the dust content in the region
close to the Sun can be more important albeit variable in time. We 
also deduce that the size of the most of the dust grains during our 
observations was less than 0.1 $\umu$m.
\end{abstract}

\begin{keywords} zodiacal dust -- Sun: general -- comets: general --
(stars:) circumstellar matter -- techniques: imaging spectroscopy
\end{keywords}

\section{Introduction} 

The work of \citet{hulst47} established, after an earlier suggestion
by Grotrian \citep[as cited by][]{hulst47}, that the emission of the
solar outer corona, called F-corona after the Fraunhofer lines
composing its spectrum, is due to the scattering of the solar light by
the dust particles.

In recent decennia it became more clear that the dust component near
the Sun cannot be described as a mere extrapolation of the zodiacal
light disc with the dust grains orbiting in the plane of the ecliptics
and slowly drifting to the Sun under the Poynting-Robertson effect as
one could accept in a first approximation.

The measurements of the Doppler shifts of the Fraunhofer lines in the
spectrum of the F-corona, first done during the total eclipse of July
31, 1981 by \citet{scheglov87}, and the analysis of the derived
velocities by \citet{shest87}, indicated that, although most of the
dust particles are orbiting in the ecliptic plane, there are also
grains orbiting at high values of inclination angle $i$, implying that
the region close to the Sun receives a dust contribution from
long-period comets.  Also, the \textit{in situ} measurements of the
dust particle velocity distribution on board of HELIOS-1 showed the
existence of two distinct components in the interplanetary dust,
cometary and asteroidal \citep{grun80}.  Among more recent results,
the high frequency of the Sun-grazing comets recorded by SOHO
\citep{macqueen91} makes also to consider the cometary contribution as
important.  Furthermore, studies of extrasolar planetary and dust
systems showed the existence of Falling Evaporating Bodies, possibly
comets \citep[e.g.][]{beust98}.  This observational progress was
followed by a thorough theoretical modelling taking into account
complex composition of the dust in the solar vicinity (see
\citealp{mann00}, \citealt{koba08}).

It would have been important to obtain further measurements of the
line-on-sight (thereafter LOS) velocities of the dust near the Sun to
have a better idea of the relative contributions of the cometary and
the ecliptic disc dust components and of their possible variation in
time.  After the first work of 1981, these measurements, to the best
of our knowledge, were successfully attempted only once, during the
eclipse of July 11, 1991 \citep{aimanov95} confirming the first
results.

Here we present new higher quality measurements of the LOS velocity of
the dust in the F-corona obtained during the eclipse on March 29, 2006
\citep[see][for a preliminary report]{shest07}, and discuss their
implications.
\begin{figure*}
	\begin{center}
	\includegraphics [width=18cm] {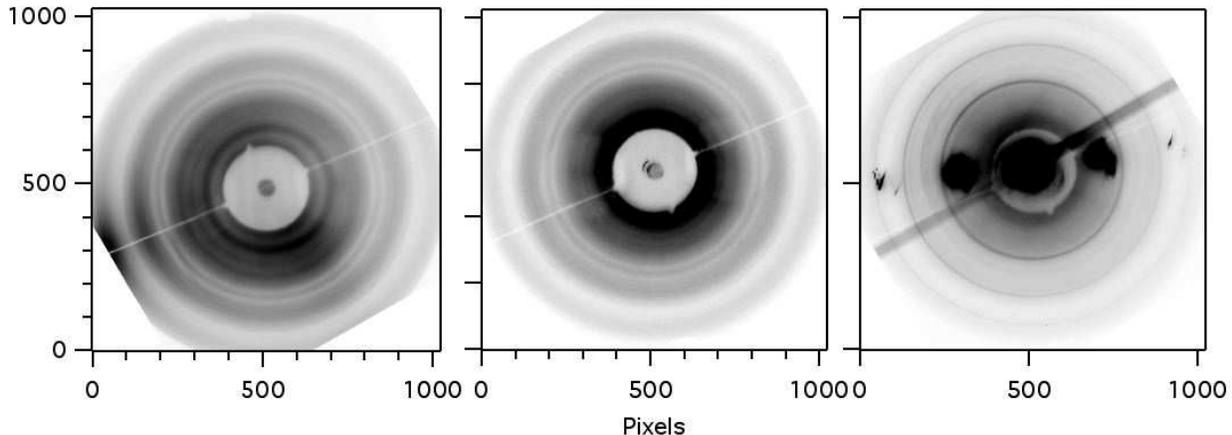}
	\end{center}
	\caption{The Fabry-P\'{e}rot interferograms of the daylight sky
	and of the circumsolar region up to elongation $\epsilon \simeq
	11$.  Here, both axes are in pixels, the platescale is 20.66
	arcsec/pix, the FOV\,$=5.9\degr$.  \textit{From left to right:} a
	comparison daylight sky frame ``D'', the total eclipse frame
	``E'', and the frame ``L'' taken at the end of the totality, but
	already including the green coronal emission line of [Fe XIV] at
	$\lambda 5302.86 \rmn{\AA}$.  The fiducial wire (black line)
	indicates the East-West direction, the West is to the left along
	the wire and the North is down.  The frames are rotated so that
	the horizontal axis of the frames coincides with the ecliptic
	plane.  The frames are dark current subtracted.  For more details
	see the text.}
	\label{fig:fig1}
\end{figure*}
\section[]{Observations}
The site of the observations was the village of Mugalzhar in the Aktobe
region of the Republic of Kazakhstan, at $\phi = 48\degr 35\arcmin$
and $\lambda = 58\degr 27\arcmin$, situated in the middle of the
totality band.  According to our calculations, the beginning of the
total eclipse was at UT 11h 32m 40, its end at 11h 35m 30s, and the
duration of the total phase was 170\,s.  The Sun was at $27\fdg 5$
above the horizon.  The weather was mostly rainy and windy during the
days preceding the eclipse, but on the eclipse day, the sky was clear
with no wind and excellent transparency.

We used the same optical set-up as in our previous measurements in
1991 \citep{scheglov87}, i.e. a Fabry-P\'{e}rot (hereafter FP)
spectrometer with a coronographic mask rejecting the light of the
solar corona and thus reducing the background.  The entrance lens of
10 cm diameter has its focal plane at the field lense, the latter is
followed by a collimator.  The FP etalon in series with an order
separation filter is placed in the parallel beam close the exit pupil
imaged by the field lense.  It is followed by a photographic objective
and a CCD detector.  The latter was an Apogee Alta-10 CCD device with
$2048^{2}$ pixels of 14 $\umu$m size.  During the data reduction, the
frames were binned by $2\times 2$ pixels, so that to avoid any
confusion we will refer from now on to the detector format of
$1024^{2}$ pixels with the 28 $\umu$m pixel size.

The solar angular radius at the time of the eclipse was
$\mathcal{R}_{\sun} = 961$ arcsec, the measured on the CCD value was
46.5\,pix.  This gives the equivalent focal length of 279.4 mm, and the
platescale of 20.66 arcsec/pixel.  The field of view (FOV) is
$5.9\degr$, corresponding to the elongations $\epsilon < 11$, however,
to reduce the background light, the central region around the Sun,
$\epsilon < 2.6$, is hidden by the coronographic mask (hereafter, the
elongation $\epsilon$ is given in angular solar radii
$\mathcal{R}_{\sun}$).
\begin{figure*} 
	\begin{center}
 	\includegraphics [width=14 cm] {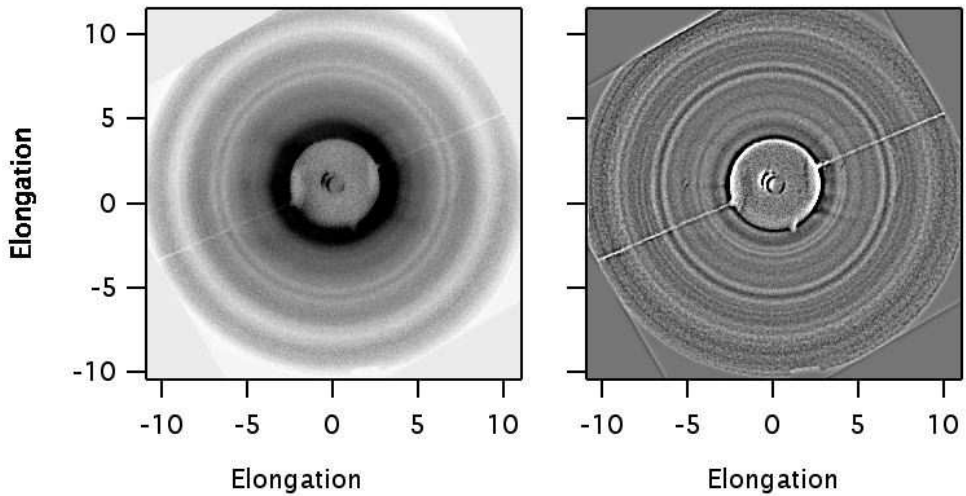}
	\end{center}
		\caption{The interferogram of the circumsolar region during
		the total eclipse in different states of data reduction
		procedure.  \textit{Left:} after median filtering and dividing
		on the flat-field as described in the text.  \textit{Right:}
		The same with the low frequency two-dimensional trend
		subtracted.  The elongation $\epsilon$ is in angular solar
		radii $\mathcal{R}_{\sun}$.}
	\label{fig:fig2}
\end{figure*}
\subsection[]{Spectro-imaging}
We used a mechanically adjustable FP etalon with the free space
$\Delta=70\,\umu$m.  The order separation filter has the FWHM =
$10\,\rmn{\rmn{\AA}}$ with the central wavelength close to that of the
\mbox{Mg\,{\sc i}} line at 5172.69 $\rmn{\AA}$.  In the wavelength
space, the set-up is such that a frame includes 4 FP orders.  The
highest, on-the-axis order $N=\Delta/\lambda$, is at the centre of the
frame and is unseen due to the use of the coronographic mask; the
orders lower than $N-4$ are outside of the FOV.

The dark subtracted frames obtained during the eclipse are shown in
Fig.\,\ref{fig:fig1}.  The left frame, denoted as ``D'', is one of the
calibration interferograms of the daylight sky, scattered on a white
screen, recorded shortly before and after the total phase of the
eclipse.  The use of the white screen allows a homogenous brightness
distribution over the field of view without changing the spectrometer
position which remains at the same position as during the total phase.
The frame shows concentric rings of the Fraunhofer absorption lines
scattered in the terrestrial atmosphere.

The dark circle at the centre of the frame corresponds to the
coronographic mask.  The dark line corresponds to a fiducial wire
indicating the East-West axis, the West to the left along the wire and
the North down.  For convenience of interpretation, the frames are
rotated so that the frames horizontal axis coincides with the ecliptic
plane.  The frames are dark current subtracted.

The frame ``E''in the middle of the Fig.\,\ref{fig:fig1} is our main
scientific exposure, of 130\,s, taken on the circumsolar region during
the eclipse.  It was started a few seconds after the beginning of the
totality and stopped well before its end, so that contamination by
coronal or chromospheric lines was avoided.  The concentric dark rings
are the solar Fraunhofer lines scattered on the dust particles of the
F-corona.  

Finally, the frame ``L''at the right of the Fig.\,\ref{fig:fig1}, with
the exposure time of 20\,s was started close to the end of the total
phase and lasted a few seconds beyond, so that it caught the light of
the solar corona which was just appearing from behind the mask (but
not yet the photosphere).  On this frame, one can see, additionally to
the absorption spectral features of the F-corona, the bright emission
rings of the green coronal [Fe XIV] line at $\lambda 5302.86
\rmn{\AA}$.  Its wavelength lies far from the centre of the our
narrow-band filter, however the emission is so strong that its light
passed in the filter transmission wings.  The corresponding
interference rings well visible in Fig.\,\ref{fig:fig1} traced the
location of the 3 used Fabry-P\'{e}rot orders and gave useful
reference points for the data reduction.  Its measured $\rmn{FWHM} =
1.2\pm 0.1\rmn{\AA}$ gives the spectral resolution of the instrument.

The daylight sky interferograms ``D'' provided the wavelength standard
and allowed to eventually measure the Doppler shift of the dust
particles of the F-corona, while the interferogram of the coronal
green line $\lambda 5302.86\rmn{\AA}$ allowed to calibrate the
spectral geometry of the frames, and in particular to accurately
define the rings centre as it will be discussed in
Section\,\ref{Section:Reduction}.

\begin{figure*} 
	\begin{center}
 	\includegraphics [width=12 cm] {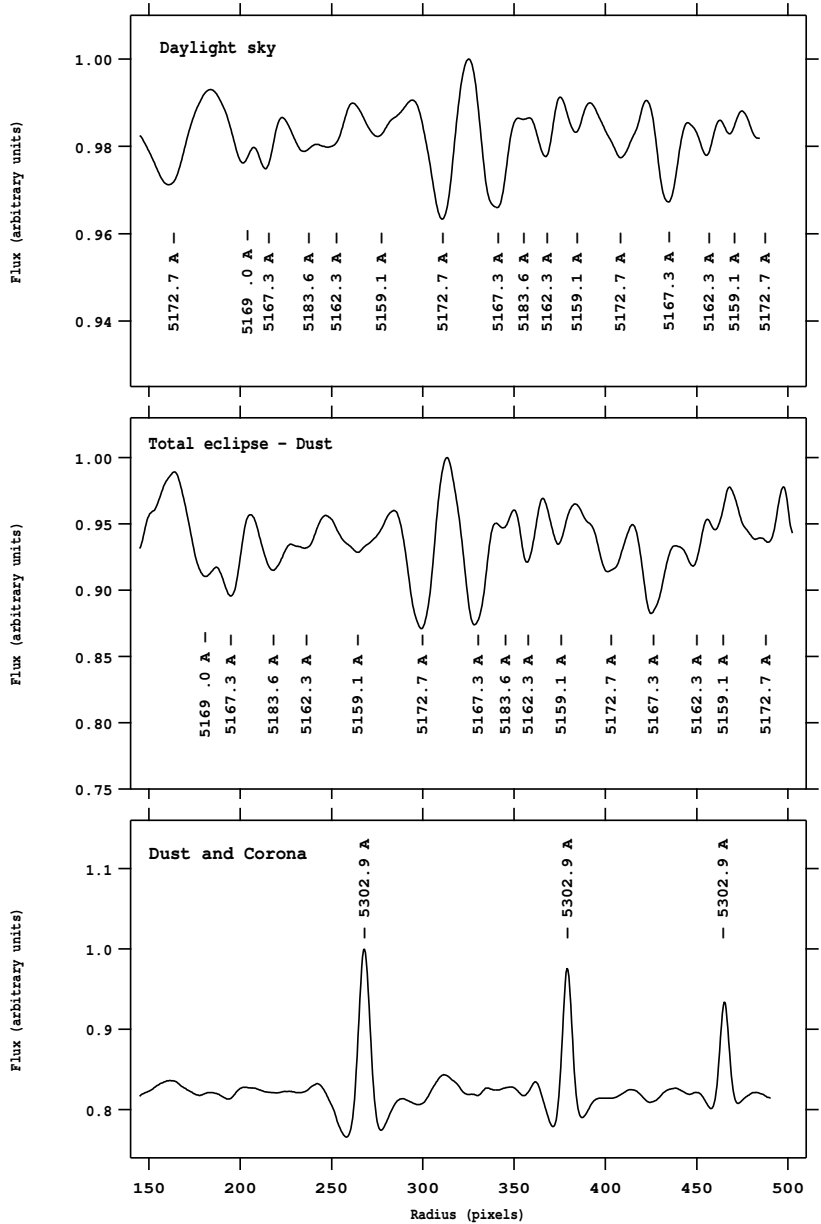}
	\end{center}
		\caption{Extracted spectrograms: The flux integrated over the
		position angle $PA$ in a function of the radius r.
		\textit{Upper plot:} The daylight sky, frame ``D";
		\textit{Middle plot:} The circumsolar region during the total
		eclipse, frame ``E"; \textit{Lower plot:} Same but also
		including the strong coronal emission [$\mbox{Fe\,{\sc
		xiv}}]\,\lambda 5302.86 \rmn{\AA}$, frame ``L".  The relevant
		Fraunhofer lines and their wavelengths are indicated.  Note
		that there are 3 overlapping spectral orders, they are traced
		by the [$\mbox{Fe\,{\sc xiv}}$] emission pics.}
	\label{fig:fig3}
\end{figure*}
\subsection[]{Brightness calibration}
The weather conditions being excellent and stable, the Fabry-P\'{e}rot
data were calibrated in brightness relative to that of the solar disc,
$B_{\sun}$.  Two 0.1\,s exposures on the Sun through a combination of
a neutral and green filters were taken to the East, at $\epsilon =
4.0$, and to the West, at $\epsilon = 4.4$ from the Sun position at
the total phase of the eclipse.  The Sun elevation for the calibration
and for the eclipse frames was nearly the same, about $30\degr$.  The
uncovered surface of the solar disc was 0.94 and 0.25 respectively for
the two exposures.  Taking this into account, the daylight sky
brightness off the eclipse is estimated to has been $4\cdot
10^{-5}B_{\sun}$.

The brightness of the F-corona continuum emission, $B_{F}$, was
measured to be $4.1\cdot 10^{-9}B_{\sun}$ at the elongation $\epsilon
= 4.0$, it was $3.9\cdot 10^{-9}B_{\sun}$ at $\epsilon = 4.4$, and
$1.7\cdot 10^{-9}B_{\sun}$ for the $\epsilon$ range from 9 to 10.  The
brightness decrement in the F-corona is in agreement with the results
of \citet{koutchmy78}.  Compared to the daylight sky, the F-corona is
fainter by a factor of $10^{4}$.
\section[]{Data reduction}
	\label{Section:Reduction}
The extraction of the Doppler shift of the solar absorption lines out
of the Fabry-P\'{e}rot interferograms needs a thorough data reduction
process.  It is worthwhile to be describe here in details.
\subsection[]{Noise and trends handling }

After the bias and dark current subtraction, all the frames were
$2\times 2$ binned, which increased the signal-to-noise ratio without
changing spectral or spatial resolution.  The next step was defining
all regions meaningless for further reduction, which includes the
central dark region corresponding to the coronographic mask, that of
the fiducial wire, the part lying out of the field-of-view of the
detector and the regions of strong light induced by spurious
reflections.  The resulting numerically masked area was about 30\%.

The ``hot", ``cold" and ``dead" pixels were cleaned out using a median
filter.  The visual analysis showed that the ``clouds" of defective
pixels did not exceed a region of 10-12 pixels a size.  We used
therefore a non-linear circular filter covering 31 neighbour pixels
(i.e. 2n+1, where n is the maximum scale of ``clouds").  The correction
was applied only to the pixels with the count exceeding 20\% of the
median value.  The number of corrected pixels was less than 4\% on the
eclipse frames and less than 0.5\% on the calibration frames.

For convenience of interpretation, all frames were then rotated so
that the horizontal axis coincides with the ecliptic plane.

The next step was the flat-fielding.  Usually, it is done by a
straightforward division by a averaged flat-field frame with a
subsequent masking of the pixels resulting from the division by zero
or close to zero values.  We decided to optimize this operation by
adding a small constant to the flat-field frame.  This is similar in
its spirit to the Tikhonov regularisation of ill-defined inverse
problems.  The value of the constant was defined by analysing the
histogram of the counts of the flat-field frame.  The aim of this
operation is to keep moderate, or negligible, the change induced in
the signal-to-noise ratio.  Thus, the division on the flat-field does
not change the structure of the frames and avoids a useless addition
of a noise.

Further, we measured the two-dimensional low frequency trend of the
resulting frames in order to subtract it and to keep only useful
spectro-spatial variations.  It was done by using iteratively a linear
circular moving average filter.  The best filter diameter and the
number of iterations were searched by trials in such a way that the
frame with the subtracted 2D trend would still keep the structures on
a 10-15 pixels scale, which is that of the recorded Fraunhofer lines.
The best filter had the diameter of 13 pixels, covering 137 closest
pixels, and the best number of iterations was 4.  Higher the number of
iterations, closer the iterative filter to a linear gaussian smoothing
filter, $exp[-(x{^2} + y{^2}/(2 \sigma_{g}^{2})]$, where $\sigma_{g}$
defines the degree of the smoothing.  The advantage of using an
iterative filter is the possibility to control the achieved smoothness
by limiting the number of iterations.

We tried square and rectangular moving average filters, however they
give rise to spur line-like features.  We also applied the methods of
high and low enveloping curves, but it did not give better results.

The FP interferogram after median filtering and subtraction of the 2D 
low frequency trend is shown in Fig.\,\ref{fig:fig2}.

Finally, the data were passed through the gaussian filter with
$\sigma_{g} = 1.3$ pixels, or $FWHM\approx 3$ pixels which is close to
the measured $FWHM$ of the point-spread function of the experiment.

\begin{table*}
	\centering
	\caption{Spectral lines identification.}
	\begin{tabular}{| c | c |c |c |}
		\hline
		Radius & Elongation $\epsilon$ & Wavelength & Element \\
		(pixels) &  ($\mathcal{R}_{\sun}$) & $\lambda (\rmn{\AA}$) & \\
		\hline
 		182 & 3.92 & 5168.91+5169.04 & \mbox{Fe\,{\sc i}} + \mbox{Fe\,{\sc ii}} \\		
		196 & 4.22 & 5167.33 & \mbox{Mg\,{\sc i}} \\
		220 & 4.72 & 5183.62 & \mbox{Mg\,{\sc i}} \\
		238 & 5.12 & 5162.28 & \mbox{Fe\,{\sc i}} \\
		266 & 5.72 & 5159.06 & \mbox{Fe\,{\sc i}} \\
		298 & 6.40 & 5172.69 & \mbox{Mg\,{\sc i}} \\
		328 & 7.06 & 5167.33 & \mbox{Mg\,{\sc i}} \\
		344 & 7.40 & 5183.62 & \mbox{Mg\,{\sc i}} \\
		358 & 7.68 & 5162.28 & \mbox{Fe\,{\sc i}} \\
		374 & 8.04 & 5159.06 & \mbox{Fe\,{\sc i}} \\
		402 & 8.64 & 5172.69 & \mbox{Mg\,{\sc i}} \\
		425 & 9.14 & 5167.33 & \mbox{Mg\,{\sc i}} \\
		448 & 9.64 & 5162.28 & \mbox{Fe\,{\sc i}} \\
		460 & 9.88 & 5159.06 & \mbox{Fe\,{\sc i}} \\
		485 & 10.42 & 5172.69 & \mbox{Mg\,{\sc i}} \\
		\hline
	\end{tabular}
	\label{tab:tab_lines}
\end{table*}
\begin{table*}
	\centering
	\caption{Areas used for the Doppler shift measurements.}
	\begin{tabular}{| c | c |c |}
		\hline
		Elongation range & Mean $\epsilon$ & Lines used \\
		\hline
		3.12 - 4.09 & 3.66 & 5172.7, 5169.0 \\
		3.12 - 4.82 & 4.09 & 5169.0, 5167.3 \\
		4.09 - 4.82 & 4.52 & 5167.3, 5183.6 \\
		4.82 - 5.63 & 5.27 & 5183.6, 5167.3 \\
		5.63 - 6.37 & 6.02 & 5159.1, 5172.7 \\
		6.36 - 7.01 & 6.71 & 5172.7, 5167.3 \\
		7.01 - 7.74 & 7.40  & 5167.3, 5183.6, 5162.3 \\
		7.74 - 8.43 & 8.08 & 5162.3, 5159.1 \\
		8.43 - 9.12 & 8.77 & 5172.7, 5167.3 \\
		9.12 - 9.63 & 9.38 & 5167.3, 5162.3 \\
		9.63 - 10.25 & 9.98 & 5162.3, 5159.1, 5172.7 \\
		\hline
	\end{tabular}
	\label{tab:tab_range}
\end{table*}

\subsection[]{Defining the interferograms centre}
In an ideal Fabry-P\'{e}rot interferogram, the wavelength $\lambda$ is
constant on a circular ring, and varies with its radius as $r{^2}$.
Let $X_{c}$ and $Y_{c}$ denote the coordinates of the centre of the
interference rings.  Their values depend on many optical parameters,
which can vary with temperature and mechanical flexures, so that they
must be carefully defined for each frame.  After different trials, we
found that the most reliable way to measure the centre position was to
use a correlation method in the following way.

Let us denote $z=r^{2}$.  We adopt a first guess of the centre
coordinates $X_{0}$ and $Y_{0}$, and divide the frame on two
sub-frames, left and right, symmetrically with the respect to $X_{0}$.
For each sub-frame, we compute counts \textit{vs} $z$, which gives us
two functions $L(z)$ and $R(z)$ respectively for the left and the
right sub-frames.  We compute then the correlation of $L$ and $R$ ,
and vary $X_{0}$.  The maximum of the correlation gives the value of
$X_{0}$, which is adopted as $X_{c}$.  In a similar way, we define the
best value of $Y_{c}$, correlating the upper and the lower sub-frames.

Such a correlation measurement can give false and biased results if,
for example, there is a strong asymmetry in the intensity of
interference patterns.  To have an additional check, we applied
another method to smaller parts of the patterns, using only arcs of
the rings.  For a given ring, we take first guess values of $X_{c}$,
$Y_{c}$, $r$, and the width of the ring $dr$.  Varying the values of
$X_{c}$, $Y_{c}$ and $r$, we find the values such that the sum of
counts is the less (for absorption lines).  To insure a good
statistics, the value of $dr$ must be sufficiently large, but without
covering neighbor ring patterns.  We used values of $dr$ in the range
from 2 to 12 pixels.  The difference of the centre coordinates defined
by this method and that of correlations did not exceed 0.3 pix, which
is at the level of the expected uncertainty.

The resulting values of the center coordinates are defined with an
accuracy better than 0.3 pix.
\subsection[]{Reduced spectrograms}
Once the centre of the interference rings was found, it is convenient
to transform the data presentation from cartesian to polar coordinates
$(r, \phi)$.  The frames were oriented so that the polar angle $\phi$
and the position angle on the sky, $PA$ are the same for all of them.
The extracted spectrograms in the form of the flux integrated over all
values of the position angle $PA$ in a function of the radius counted
from the interferogram centre are plotted in Fig.\,\ref{fig:fig3} for
the daylight sky, the circumsolar region during the total eclipse and
the frame ``L" including the coronal $\mbox{Fe\,{\sc xiv}}, \lambda
5302.86 \rmn{\AA}$ line.  The 3 pics of the coronal emission line
trace the 3 Fabry-P\'{e}rot spectral orders.  The relevant scattered
$\mbox{Fe\,{\sc i}}$, $\mbox{Fe\,{\sc ii}}$ and $\mbox{Mg\,{\sc i}}$
Fraunhofer lines, in absorption, are indicated.  Their wavelengths,
the values of the ring radius $r$ and the corresponding value of the
elongation $\epsilon$ are given in the Table\,\ref{tab:tab_lines}.
Albeit the used Fabry-P\'{e}rot orders overlap, the spectral features,
fortunately, are distinct and can be easily identified.

\begin{figure*} 
	\begin{center}
 	\includegraphics [width=12 cm] {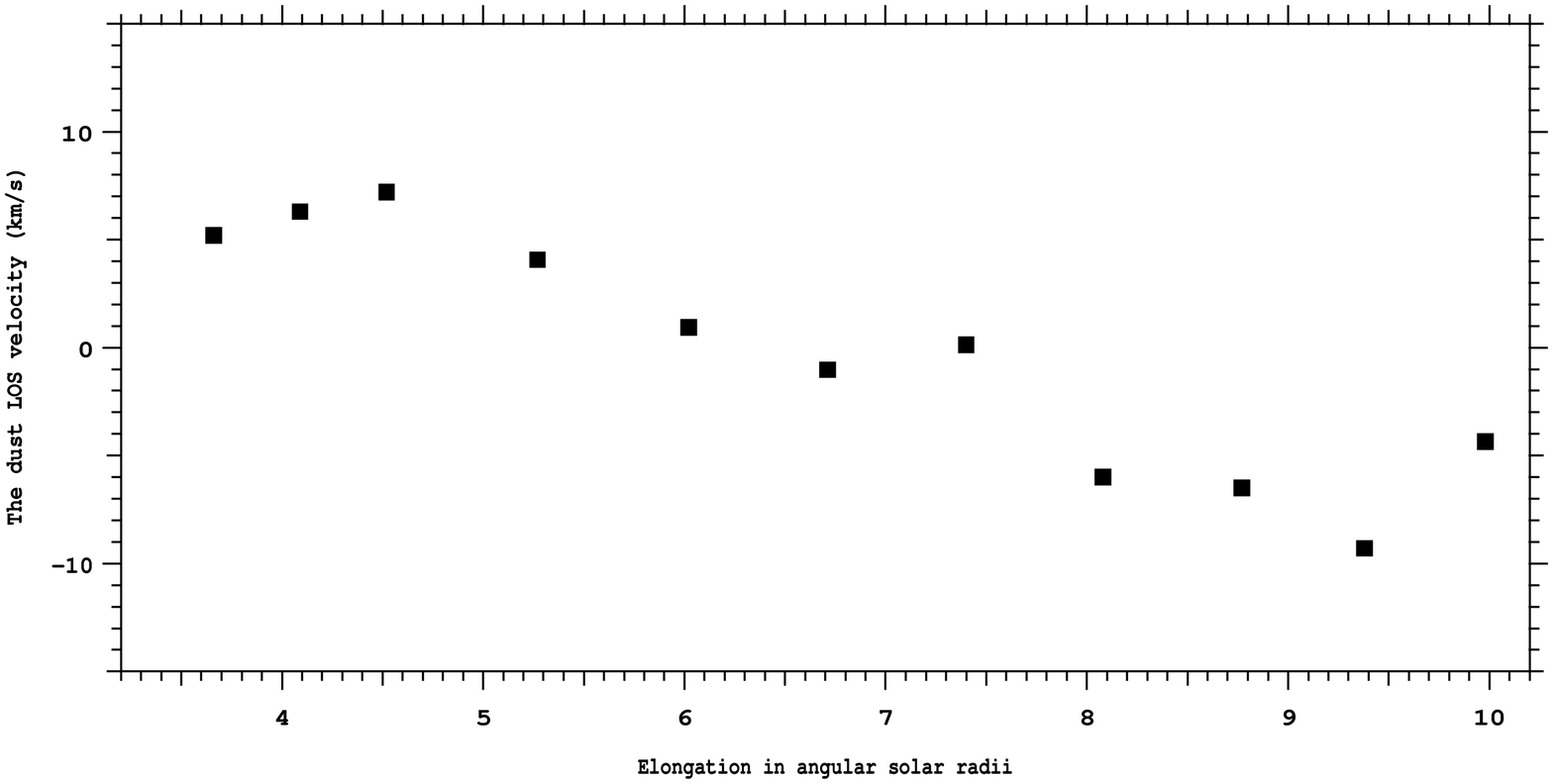}
	\end{center}
		 \caption{The line-on-sight velocity of the dust
		 $\bar{V}_{\epsilon}$ \textit{vs} the elongation $\epsilon$;
		 $\bar{V}_{\epsilon}$ is the average over all range of the
		 position angle $PA$.}
	\label{fig:fig_vel_elong}
\end{figure*}
\begin{figure*} 
	\begin{center}
 	\includegraphics [width=12 cm] {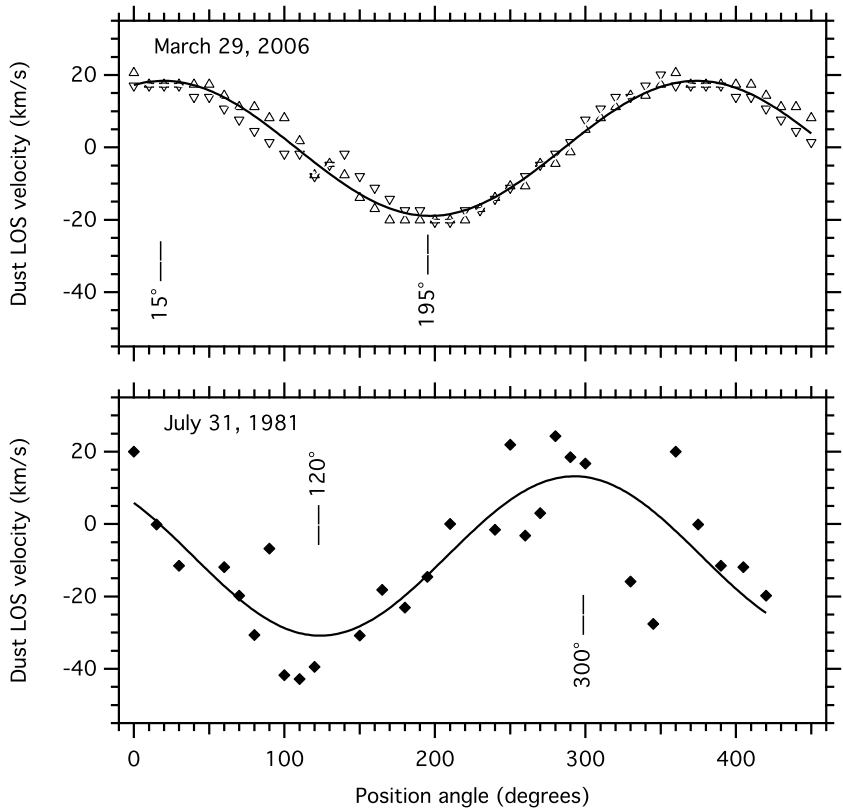}
	\end{center}
		\caption{The line-on-sight velocity of the dust averaged over
		the studied range of elongations, $\bar{V}_{PA}$, \textit{vs}
		the position angle $PA$ on March 29, 2006 (upper plot) and on
		July 31, 1981 (lower plot) together with the fitted $sinus$
		curves.  For 2006, the triangle marks indicate measurements
		using two different daylight sky interferograms.  The values
		for the $PA$ range $360\degr-450\degr$ were added for
		convenience, they merely repeat those for $0-90\degr$.  The
		$PA$ values of the extrema of the fitted to $V_{d}$ $sinus$
		curves are given.  For the circular orbits lying strictly in
		the ecliptic plane, the extrema should be at $90\degr$ and
		$270\degr$.}
	\label{fig:fig_vel_pa}
\end{figure*}
\begin{figure*} 
	\begin{center}
 	\includegraphics [width=12 cm] {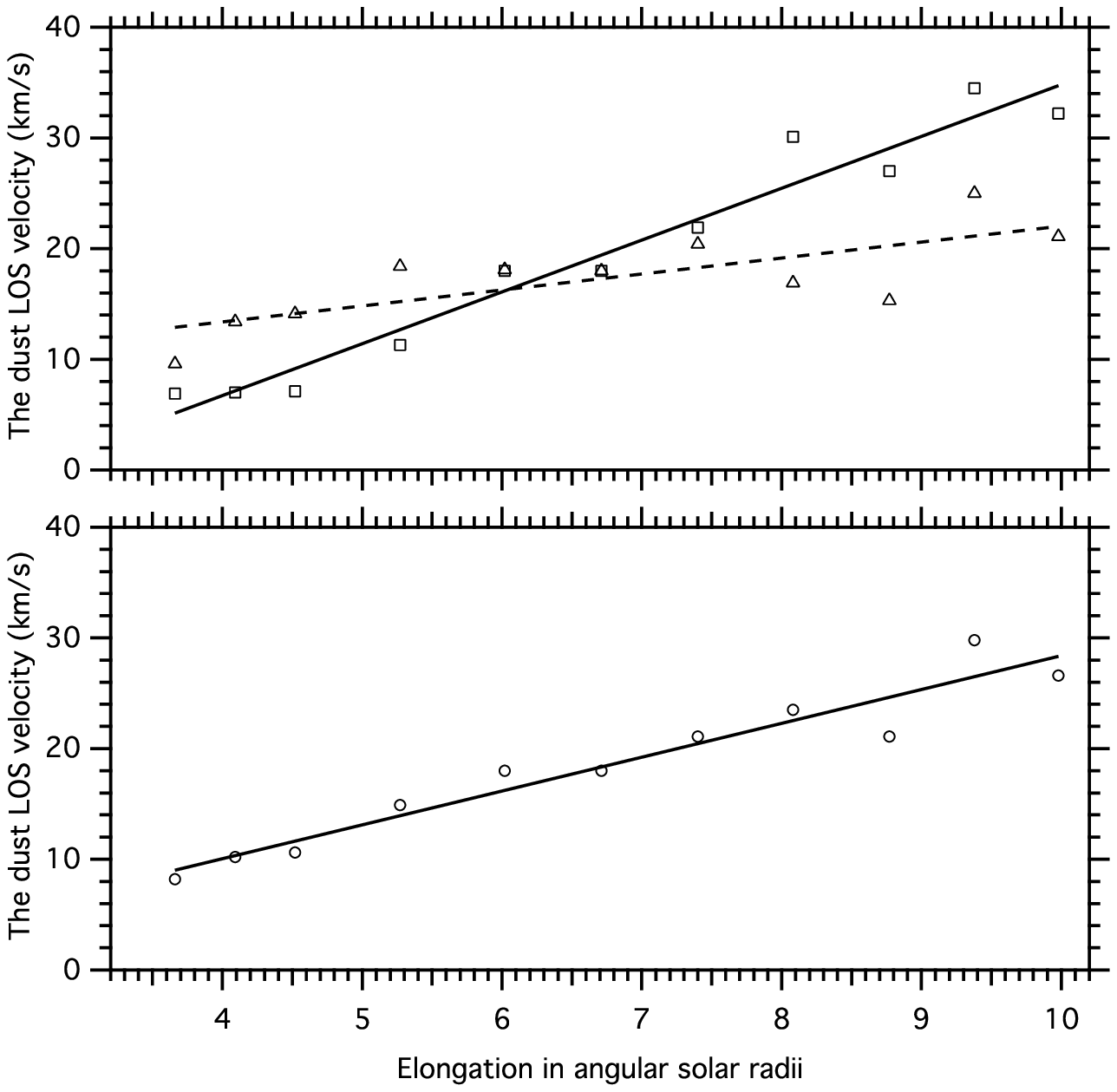}
	\end{center}
		\caption{\textit{Upper plot}: The difference $V_{orb} = V -
		\bar{V}_{\epsilon}$ \textit{vs} the elongation $\epsilon$ in
		the directions of $PA=15\degr$ (rectangles) and $PA=195\degr$
		(triangles) and the corresponding linear fits.  The values of
		$V$ are the average over $\pm30\degr$ wide sectors, they are
		given in absolute values.  The actual sign is positive in the
		first direction and negative in the second one.  \textit{Lower
		plot}: The average of the two and the linear fit to it,
		$V_{orb} = -2.2 + 3.1 \epsilon $}
	\label{fig:fig_vel_orb}
\end{figure*}
\subsection[]{The line-on-sight velocity of the dust.  }
The Doppler shift between the eclipse and the daylight spectrograms
was measured by cross-correlation.  The corresponding dust LOS
velocity $V$ writes:
 \begin{equation}
	V = \frac{c \Delta_{r}^{2}}{2f{^2}}
 	\label{eq:eq1}
 \end{equation}
or, substituting the numerical values:
 \begin{equation}
	V= \frac{3\times 10^{6}\times 0.028{^2}\times
	\Delta_{r}^2}{2\times 86.65{^2}} = 0.01566 \times \Delta_{r}^2
 	\label{eq:eq2}
 \end{equation}
where $f = 86.55$ mm is the focal distance of the camera objective,
0.028 mm is the pixel size, $\Delta_{r}^2 = r_{e}^{2} - r_{sky}^{2}$,
where $r_{e}$ and $r_{sky}$ are the radii in pixels respectively for
the eclipse and daylight sky frames.  The pixels convert to the
elongation $\epsilon$ given that $\mathcal{R}_{\sun} = 46.5$ pixels.

The sampling in elongation $\epsilon$ and in position angle $PA$ was
as follows.  We used intervals of $\epsilon$ as given in
Tab.\,\ref{tab:tab_range} choosing their spanning as regular as
possible in a function of the useful spectral lines.  The full range
of $PA$ was simply divided on 36 sectors $20\degr$ wide each.

For convenience of the analysis, we computed two kinds of average
velocities, $\bar{V}_{\epsilon}$, which is the average over all values
of $PA$, and $\bar{V}_{PA}$, which is the average over all values of
the elongation $\epsilon$.  

The  value $\bar{V}_{\epsilon}$ is the function of $\epsilon$
only; it is plotted in Fig.\,\ref{fig:fig_vel_elong}.  If the LOS
velocities of the dust grains are distributed in a central symmetry
with the respect to the Sun, as it is expected for a pure circular
motion, then $\bar{V}_{\epsilon}$ is 0; if not, it reflects the
presence of a radial motion, an infall or outflow, at this particular
value of $\epsilon$.  The Fig.\,\ref{fig:fig_vel_elong} shows that
radial motions with velocities about $10\,\rmn{km}\cdot\rmn{s}^{-1}$
or less might be present at $\epsilon=4.5$ and $\epsilon\approx9$, and
they are absent, or very small, in between these elongations.

The value $\bar{V}_{PA}$ is a function of the position angle $PA$
only, it is plotted in Fig.\,\ref{fig:fig_vel_pa}.  It is so
particular that we give also for comparison a similar plot from the
eclipse on July 31, 1981.  For 2006, the triangle marks indicate
measurements using two different daylight sky interferograms; their
scattering provides an estimate of the uncertainty $\delta
\bar{V}_{PA}\simeq 1.6\,\rmn{km}\cdot\rmn{s}^{-1}$.  The values for
the $PA$ range $360\degr-450\degr$ are added for convenience, they
merely repeat those for $0-90\degr$.  Plotted are also the least
squares $sinus$ fits in the form: $\bar{V}_{PA} = k_{1} + k_{2}\cdot
sin(k_{3}\cdot PA +k_{4})$.  The fit coefficients, for 2006, are as
follows: $k_{1} = 0.3\pm 0.3$, $k_{2} = 18.7\pm 0.3$, $k_{3} =
0.01773\pm 0.00016$ and $k_{4} = 1.22\pm 0.04$, and for 1981, $k_{1} =
8.8\pm 2.5$, $k_{2} = 22.0\pm 3.5$, $k_{3} = 0.019\pm 0.001$ and
$k_{4} = 2.4\pm 0.3$.
\section{Discussion}
	\label{Section:Discussion}
	
\begin{table*}
	\centering
	\caption{Orbital elements of sungrazing comets at the end of
	March, 2006}
	\begin{tabular}{| c | c |c | c | c |c |}
		\hline
		Comet name & Epoch & Perihelion q  & Perihelion  & 
		Longitude  of & Inclination i \\
		& &  in $AU$ &  argument & the node & \\
		\hline
		CK06F050 & March 21.96 & 0.0050 &  82.38 & 4.20 & 144.57 \\
		CK06F060 & March 23.04 & 0.0333 &  56.09 & 75.03 & 74.13 \\
		CK06F070 & March 28.64 & 0.0050 &  84.83 & 3.42 & 145.72 \\
		CK06F080 & March 31.10 & 0.0052 &  84.02 & 5.67 & 144.58 \\
		\hline
	\end{tabular}
	\label{tab:tab_comets}
\end{table*}

For the prograde orbits being strictly in the ecliptic plane, the
minimum LOS velocity $\bar{V}_{PA}$ should be at $90\degr$, i.e.\,to
the East from the Sun, and the maximum $\bar{V}_{PA}$ at $270\degr$,
i.e.\,to the West.  The velocity curve measured on July 31, 1981, is
close to what is expected from such a motion (with, possibly, a slight
difference, the minimum of $\bar{V}_{PA}$ being at $120\degr$ and the
maximum at $300\degr$).
 
As to the $\bar{V}_{PA}$ curve on March 29, 2006, the values of $PA$
of the extrema differ dramatically from what is expected from the
prograde motion in the ecliptics, namely the minimum $\bar{V}_{PA}$ is
at $195\degr,$ and the maximum is at $15\degr$, meaning a retrograde
motion in a plane nearly perpendicular to the ecliptics.

The dust rotating in the prograde direction in the ecliptics is also
barely present showing the "jump" of $\bar{V}_{PA}$ at $PA \simeq
120\degr$ and the symmetric to it another "jump" at $PA \simeq
300\degr$ (see Fig.\,\ref{fig:fig_vel_pa}, upper plot).

Let us verify whether the the LOS velocity behavior agrees with what
one would expect from the keplerian motion.  This can be seen from the
variation of the LOS velocity with the elongation $\epsilon$ at a
given $PA$.  We assume that the elongation $\epsilon$ of the
scattering dust and its heliocentric distance are in a linear
relation.  We choose the values of $15\degr$ and $195\degr$ for the
$PA$, which corresponds to the extrema of $\bar{V}_{PA}$ curve.  First
of all, the LOS velocities in the considered $PA$'s were averaged
within a sector of $\pm30\degr$ wide, then we subtracted from it the
$\bar{V}_{\epsilon}$ value.  The resulting ``orbital" LOS velocity
$V_{orb}(\epsilon )$ is given in Fig.\,\ref{fig:fig_vel_elong}
together with the linear fit.  The latter gives
$\bar{V}_{orb}(\epsilon) = -2.2 + 3.1\epsilon$ with the $1\sigma$
uncertainty of $\pm2$ on the additive term and $\pm0.3$ on the factor.
The success of the fit indicates that the motion is indeed keplerian.
	
Summarizing what is indicated by the $\bar{V}_{PA}$ curve on March 29,
2006, we conclude that the bulk of the dust grains was in a keplerian
retrograde motion in a plane nearly perpendicular to the ecliptics.
In the inner Solar system, the only objects having this kind of
motions are comets.

It happens that indeed around the eclipse date, the SOHO spacecraft
recorded a series of sungrazing comets \citep{soho06} of the Kreutz
group (\citealt{marsden67}, \citealt{sekanina07}).  One of them,
CK06F050, fell onto the Sun a day before the eclipse, another one,
CK06F070, fell a day and a half after the eclipse.  For convenience of
the reader, the orbital elements as given by \citet{soho06}, are
reproduced in the Tab.\,\ref{tab:tab_comets}.  It is quite possible
that there were also a chain of smaller fragments in between the
entities recorded by SOHO, and thus the FOV of our eclipse frame is
filled by the dust brought by the Kreutz group comets moving to the
Sun on retrograde orbits.

The date of our observations is 8 days past the equinox, so that the
line-of-sight to the Sun projects to approximately 8$\degr$ East from
that to the vernal point $\Omega$.  The longitude of the node of the
indicated comets is distant from the $\Omega$ point, according to
Tab.\,\ref{tab:tab_comets}, by $3\degr-6\degr$.  This means that our
line-of-sight is inclined with the respect to the line of nodes by
only $2\degr - 5\degr$.  The perihelion argument lies at
$84\degr-85\degr$ from the line of nodes.

This means that our image plane is nearly perpendicular to the orbital
plane of the comets, and the line-of-sight slides over it.  

If the dust grains were strictly confined to the orbital plane, we
would see the selected direction indicated by the extrema of the LOS
velocities at $PA\approx55\degr$, which would correspond to the
inclination of the parent comet's orbit of $i = 145\degr$.  But our
data indicate a different axis for the extrema velocities, namely
$PA\approx15\degr$, which indicates $i = 105\degr$, i.e.\,the orbital
plane of the dust grains is turned with respect to that of the comets
by $40\degr$.  The orbital plane cannot change under the gravitational
force.  Hence, we have to assume another reasons, the effect of the
magnetic field on electrically charged dust particles being the most
plausible.  It follows that the size of particles is quite small,
which, in turn, suggests their cometary origin.  According to the
detailed models of the dust grains dynamics near the Sun by
\citep{krivov98} and \citep{mann00}, the Lorentz force dominates the
gravity for the dust grains smaller than 0.1 $\umu$m i size, and, as
show their numerical simulations \citet{mann00}, the orbits of such
grains get randomized which is not th case for the grains with size
larger than 0.1 $\umu$m.  Interestingly, for grains with the size
smaller than 0.01 $\umu$m, the ratio charge-to-mass may be so high
that they can be accelerated outward by the interaction with the solar
wind as shows their detection by STEREO experiment \citet{meyer09}.

A more detailed modelling of the presented data, taking into account
the scattering geometry, dynamics, the density and size distributions
of the dust grains, would be highly desirable to quantify further the
dust physical properties near the Sun, but this is largely beyond the
scope of the present article.
\section{Conclusions}
The reported here measurements of the Doppler shifts of the Fraunhofer
lines, scattered by the dust grains in the solar F-corona, show that
at the date of our observations the dust grains were on the orbit with
a retrograde motion in a plane at $i\approx 105\degr$, i.e. nearly
perpendicular to the ecliptics.  This points to their cometary origin.
Indeed, at the end of March, 2006, SOHO recorded several sungrazing
comets with the orbital elements close to what was deduced from our
measurements.  We conclude that the contribution of comets to the dust
content in the region close to the Sun can be important albeit
variable in time.  This contribution can explain the already noticed
change of the dust distribution from one in the axial symmetry far
from the Sun to that in the central, or may be spherical, symmetry
\citet{mann00}.

We also derive that the observed plane of the dust grains orbit is
slightly different from that of the parent comet(s), which indicates
that the size of grains is small, less than 0.1 $\umu$m, so that they
are deviated from the initial orbit by the Lorentz force. This also 
means that the observed dust grains were released by the comet(s) 
shortly before our observations.

The importance of comets in a circumstellar environment is general,
let us recall e.g. the ``Falling Evaporating Bodies'' recorded in the
spectra of $\beta$ Pic \citep[e.g.][]{beust98}.  It would be
interesting to investigate whether they can provide a sufficient
transport of the dust grains between far and close environments of the
central star, and to contribute to the dissemination of crystallized
material in the recently detected exozodiacal dust discs
(\citealp{absil06}, \citeyear{absil09}).  For a more detailed
discussion on the possible role of comet see also \cite{augereau09}
and references therein.
\section {Acknowledgments}
We are indebted to many persons who helped to make these measurements
possible.  A.\,Dubovitskiy, G.\,Minasiantz, M.\,Bayiliev insured the
transportation of the instrument and helped with its set-up at the
site of the eclipse.  A.\,Didenko provided the Apogee CCD camera,
T.\,Hua supplied a high quality filter on the \mbox{Mg\,{\sc i}} line
used in this work.  We benefited from useful discussions with E.\,le
Coarer, T.\,Bonev, V.\,Golev, A.-M.\,Lagrange, J.-C.\,Augereau.  The
expedition to the site of the solar eclipse was funded by the Kazakh
National Space Agency and Laboratoire d'Astrophysique de Grenoble (UMR
5571 of the CNRS and Universit\'{e} Joseph-Fourier), the work on the
data analysis became possible thanks to the EGIDE administrated
ECO-NET grant 18837\,-\,XG of the French Ministry of Foreign Affaires.


\bibliography{eclipse06} 
\bibliographystyle{aa} 

\bsp

\label{lastpage}

\end{document}